# Automatic phase space generation for Monte Carlo calculations of Intensity Modulated Particle Therapy

Qianxia Wang[1, 2*], Cong Zhu[2, 3], Xuemin Bai[2, 4], Yu Deng[5], Nicki Schlegel[5], Antony Adair[1, 2], Zhi Chen[5], Yongqiang Li[5], Michael Moyers[5], Pablo Yepes[1, 2*]

[1] Department of Physics and Astronomy, MS 315, Rice University, 6100 Main Street, Houston, TX 77005, United States of America, [2] Department of Radiation Physics, Unit 1420, The University of Texas MD Anderson Cancer Center, 1515 Holcombe Blvd., Houston, TX 77030, United States of America, [3] University of Texas Health Science Center at Houston, School of Public Health, Houston, TX, United States of America, [4] Department of Industrial Engineering, University of Houston, Houston, TX, United States of America, [5] Shanghai Proton and Heavy Ion Center, Shanghai 201315, People's Republic of China.

E-mail: qw14@rice.edu and yepes@rice.edu

Key words: automatic algorithm, phase space, Monte Carlo, particle therapy, measurement, dose calculation

Abstract: Monte Carlo (MC) is generally considered as the most accurate dose calculation tool for particle therapy. However, a proper description of the beam particles kinematics is a necessary input for a realistic simulation. Such a description can be stored in phase space (PS) files for different beam energies. A PS file contains kinetic information such as energies, positions and travelling directions for particles traversing a plane perpendicular to the beam direction. The accuracy of PS files play a critical role in the performance of the MC method for dose calculations. A PS file can be generated with a set of parameters describing analytically the beam kinematics. However, determining such parameters can be tedious and time consuming. Thus, we have developed an algorithm to obtain those parameters automatically and efficiently. In this paper, we present such an algorithm and compare dose calculations using PS automatically generated for the Shanghai Proton and Heavy Ion Center (SPHIC) with measurements. The gamma-index for comparing calculated depth dose distributions (DDD) with measurements are above 96.0% with criterion 0.6%/0.6 mm. For each single energy, the mean difference percentage between calculated lateral spot sizes at 5 different depths and measurements are below 3.5%.

## 1 Introduction

The number of patients treated with particle therapy (PP) (Wilson 1946, Amaldi and Kraft 2005) has significantly increased in recent years (PTCOG website). Compared with traditional photon therapy, the most important advantage for PP is its improved tumor coverage, while minimizing toxicity to healthy tissues (Castro *et al* 2004, Schulz-Ertner and Tsujii 2007, Ohno 2013, Poludniowski *et al* 2015). Until recently, most treatment planning systems (TPS) used analytical methods as the dose calculation tool. Such methods are typically based upon pencil beam algorithms (Petti

1992, Russell *et al* 1995, Hong *et al* 1996, Deasy 1998, Schneider *et al* 1998, Schaffner *et al* 1999, Szymanowski and Oelfke 2002, Taylor *et al* 2017) which perform well in homogeneous media, however they become less accurate when significant inhomogeneities are present, for instance in patients with tumors in the thoracic, head-and-neck, or lung regions (Sawakuchi *et al* 2008, Paganetti 2012, Yang *et al* 2012, Bueno *et al* 2013, Schuemann *et al* 2014). Monte Carlo (MC) methods, like Geant4 (Agostinelli *et al* 2003, Allison *et al* 2006), Fluka (Ferrari *et al* 2005, Böhlen *et al* 2014), MCNPX (Waters et al, 2005), etc., which can precisely describe particle transport through matter, are considered the most accurate methods for radiotherapy dose calculations (Schaffner *et al* 1999, Newhauser *et al* 2008, Titt *et al* 2008, Koch *et al* 2008, Taddei *et al* 2009, Harding *et al* 2014, Giantsoudi *et al* 2015, Dedes *et al* 2015). To overcome the limitation of long computing time for traditional MC codes, a variety of fast MC methods have been developed (Jia *et al* 2012, Giantsoudi *et al* 2015). The Fast Dose Calculator (FDC), introduced by Yepes et al (Yepes *et al* 2009a, 2009b, 2010a, 2010b), is one of these algorithms. The accuracy of this fast MC method has been validated for proton and carbon therapy in previous studies (Yepes *et al* 2016a, Yepes *et al* 2016b, Wang *et al* 2018, Wang *et al* 2019).

Beam characterization is a crucial input to the MC to obtain accurate dose distributions. In principle, fully simulating the treatment head geometry (Peterson *et al* 2009, Sawakuchi *et al* 2010) to characterize the beam is considered to be the most accurate approach, because it accounts for interactions between beam and radiation head (IEC, 1996) elements. However, such an approach is computational expensive since it requires the detailed simulation of the passage of particles through the radiation head. In addition, the radiation head geometry is usually confidential to external users, and may not be fully available. In any case, a model of the beam upstream of the radiation head is needed.

An alternative approach is to model the beam downstream of the radiation head or nozzle (Grevillot *et al* 2011, Grassberger *et al* 2015, Tessonnier *et al* 2016), but upstream of patient dependent elements (i.e. range shifters). The beam description at that location is usually more complex than upstream of the radiation head. In either case, the beam models are adjusted to reproduce measured depth and lateral dose distributions close to the isocenter (where patient located). Because the contribution of secondary particles produced in the radiation head is negligible for Intensity Modulated Particle Therapy (IMPT) (Grassberger *et al* 2015), the beam at the radiation head can be described by only including primary particles (like proton or carbon).

In this work, parameterized phase space (PS) files are used to store simulated particles traversing a plane perpendicular to the beam, and located at the radiation head exit. PS files are generated for each nominal energy available at the facility and each PS file includes a sample of 1M particles. Each particle is defined by a kinetic energy, two coordinates (x,y) on the transverse plane and a direction, described by the particle direction projected onto the transverse axes (x,y). The energy distribution for each nominal energy is characterized by a Gaussian function. Both the spatial position and the angles are parametrized with Gaussian functions or functions obtained from experimental measurements. In total, six parameters are used to describe these distributions. These parameters are tuned to match simulated depth dose distributions in water and spot sizes of lateral profiles measured in air. (Note that the spot size is defined as the Full Width at Half Maximum (FWHM) of the lateral profile along the x or y axis.) This process of parameterizing the phase space is usually quite labor intense and time consuming.

E Herranz (E Herranz *et al* 2015) and D. Sarrut (D. Sarrut *et al* 2019) have introduced their automatic phase space determination algorithms for electron or photon therapy. To our knowledge, this is the first work to develop an automatic algorithm for PS parameters determination for proton or ion therapy. In section 2, the method to determine PS parameters is introduced along with the details of the calculation. The automatic algorithm is described at length in Section 3. Results of using this algorithm to generate PS files for SPHIC is presented in Section 4. The last section is the summary.

## 2 Method to determine parameters for phase space file and calculation details

### 2.1 Method of determining parameters for phase space file

A PS file contains simulated particles, for each of the energies available from the accelerator, aimed at the isocenter as they traverse a plane perpendicular to the beam axis at the radiation head exit. Each particle is described by: a kinetic energy, a position on the transverse plane (x,y), and a direction, given as the projection of the particle momentum on the x and y transverse axes ($\cos\phi_x$, $\cos\phi_y$). A Gaussian function is used to describe the energy distribution for all particles, which is determined by two variables, its average value ($E_r$) and its standard deviation ($\sigma_E$). Gaussian functions centered at zero are also used to describe the x and y position distribution for proton, and $\cos\phi_x$ and $\cos\phi_y$ for both proton and carbon. For the highest energy carbon ion beam (430.1 MeV/u) which has asymmetric lateral profiles, instead of using Gaussian functions for the carbon x and y distributions, measured lateral dose profiles are used as the input. Since these distributions are centered at zero, we only need the width for each Gaussian to characterize them. Two variables are thus used to describe x and y: $\sigma_x$ and $\sigma_y$, and two variables for the angular distributions: $\sigma_{\phi x}$ and $\sigma_{\phi y}$. For carbon ions however, $\sigma_x$ and $\sigma_y$ are defined as scaling factors for the input lateral dose profiles. $E_r$ and $\sigma_E$ will be referred to as the longitudinal parameters since they are mainly determined from the shape of the Depth Dose Distributions (DDD). $\sigma_x$, $\sigma_y$, $\sigma_{\phi x}$, $\sigma_{\phi y}$ will be referred to as the lateral parameters, since they mainly control the shape of the lateral distributions.

At SPHIC, the accelerator can deliver 290 nominal energies for proton and 291 nominal energies for carbon with 5 foci being available for each of these energies. Different foci correspond to different beam transversal dimension at the isocenter (Tessonnier *et al* 2016). The radiation head exit is at 1125.8 mm upstream from the isocenter.

Parameters are tuned for a few selected nominal energies by comparing the calculated DDD in water, and x and y lateral profiles at five positions along the beam (z-axis) in air with measured data. The process is divided into two steps (Figure 1). Firstly, longitudinal parameters are tuned by comparing the calculated DDD with measurements, while the lateral parameters are set to some default values. Once the longitudinal parameters are optimized, we then compare the lateral dose profile spot sizes to optimize the four lateral parameters. Parameters are tuned in this order because lateral profile details have little influence on DDD, and DDD is nearly independent on lateral parameters.

We assign initial values to parameters to generate a sample of $10^6$ particles in a PS file, then run FDC to obtain DDD in water or lateral dose distribution in air. In the next step, we compare calculated depth or lateral dose distributions with measured data. Parameters are tuned accordingly until simulations match measurements. We set tolerances for different quantities such as peak position, FWHM of DDD, spot sizes at different depths and the gradient of spot sizes at two different depths of lateral dose profiles. If the difference between calculation and measurement for these variables is below the corresponding tolerance, the simulation is considered to match the measurement.

**2.2 Calculation and measurement details**

The dimensions of water phantom for DDD calculations are 300x300x500 $mm^3$. We use rectangular tallies with 50 mm size in the transverse axes, and 0.2 mm size along beam direction. We used rectangular tallies instead of a 50 mm diameter cylinder, used as the sensitive area in the measuring device, because it was easier to implement in the code, and no significant difference was observed between the two tallies. We use large tally size along the perpendicular axes in order to minimize the calculation time. The tally size along z is selected to be the same as the finest resolution of the DDD measuring device.

The dimensions of the air phantom for calculating lateral dose profile spot size are 100x100x2264 $mm^3$. To save computing time, we use tally dimensions of 0.2x10x8 $mm^3$ and 10x0.2x8 $mm^3$ for x and y axes respectively, when calculating lateral profile spot sizes. We checked that relatively larger tally size on one axis did not have a significant impact in the calculated spot size on the other axis. A small tally size (0.2 mm) is necessary for the axis of interest, because the resolution of measured spot size is one tenth of a mm. The measurements in air were taken at the following distances from isocenter along the beam axis (z): $z_{-2}$=-800 mm, $z_{-1}$=-400 mm, $z_0$=0, $z_1$=200 mm and $z_2$=600 mm.

The number of histories was set to be 100M for each FDC run. However the running will be terminated early if the mean voxel statistical error reaches the set statistical threshold value. (The statistical error is averaged over all voxels with a dose higher than 10% of maximum dose.) The acceptable mean statistical error is set to be 0.2% and 0.35% for DDD calculation in water and lateral dose calculation in air, respectively. FDC was run in GPU mode using on NVIDIA GeForce GTX 1080 GPU card.

When tuning $E_r$ and $\sigma_E$ by comparing measured and calculated DDD in water, the initial value of $E_r$ is assigned to the nominal energy for both proton and carbon. The initial value for $\sigma_E$ is 2 MeV for proton and 24 MeV for carbon. The default values for $\sigma_x$, $\sigma_y$, $\sigma_{\phi x}$, $\sigma_{\phi y}$ are 1.0, 1.0, 0.05 and 0.05 for proton and 0.1, 0.1, 0.01 and 0.01 for carbon in this step. After the $E_r$ and $\sigma_E$ are determined, we tune $\sigma_x$, $\sigma_y$, $\sigma_{\phi x}$ and $\sigma_{\phi y}$ by comparing lateral dose spot sizes. The initial values for $\sigma_x$ and $\sigma_y$ are different for proton and carbon, because for protons a Gaussian is used to describe the x and y position distributions with $\sigma_x$ and $\sigma_y$ as their width however, for carbon ions the lateral dose distribution of 430.1 MeV/u focus 3 at isocenter is used as the position distribution shape, and $\sigma_x$ and $\sigma_y$ are scaling parameters to broaden or narrow the input lateral dose distribution.

The depth dose distributions were acquired in high-resolution by using a PTW Peak finder water column (PTW, Freiburg, Germany) measuring system. All lateral profile measurements were taken with a MWPC (multi-wire proportional chamber) which is constructed of separated wires in horizontal and vertical directions. These wires record accumulated charge and a fitting program calculates the width and position of an assumed Gaussian shaped profile using the charge distribution. The MWPC was located at five positions: 800 mm upstream, 400 mm upstream, isocenter, 200 mm downstream and 600 mm downstream. 27 energy levels for proton and 15 for carbon were measured at each position.

## 3 Algorithm for automatic code

### 3.1 Parameters tuning for Depth Dose distribution

When tuning parameters in the first step, $E_r$ needs to be adjusted when the DDD simulated and measured Bragg-peaks do not match. For both measured and calculated DDD, the position of the Bragg Peak (P) is determined by finding the position of the maximum in the DDD. We related the change in $E_r$ ($\Delta E_r$) to $\Delta P = P_M - P_C$, where $P_M$ and $P_C$ are the peak positions of the measured and calculated distributions. The change in the parameter $E_r$ is given by $\Delta E_r = \omega \Delta P$, where $\omega$ is given an initial value of 0.1 MeV/mm. However, this value depends on the two previous steps. If $\Delta P_1 \Delta P_2 > 0$ (with

indices1 and 2 corresponding to the two steps before the current step), it is an indication that the ω may not be large enough. In that case, ω is scaled by 1.25. On the contrary, when $\Delta P_1 \Delta P_2 < 0$, ω is divided by 1.25. The tolerance for $\Delta P$ is set to be 0.2 mm in the code.

When tuning the longitudinal parameters, $\sigma_E$ needs to be varied until the widths of the simulated and measured Bragg peaks are close enough. We evaluate such an agreement by calculating the Full With Half Maximum, FWHM, that we will refer to as $\Sigma$, for the measured ($\Sigma_M$) and simulated ($\Sigma_S$) DDD distributions, and by the difference, $\Delta\Sigma = \Sigma_M - \Sigma_S$.

- If $|\Delta\Sigma| > 0.15$: $\Sigma_M$, $\sigma_E$ is varied by 50% in each step: $\sigma_E = \sigma_E + 0.5\ \sigma_E\ \Delta\Sigma/|\Delta\Sigma|$.
- If $|\Delta\Sigma| < 0.15$: $\Sigma_M$, $\sigma_E$ is varied by $R_E\ \Delta\Sigma/\Sigma_M$ in each step: $\sigma_E = \sigma_E + R_E\ \Delta\Sigma/\Sigma_M$.

$R_E$ is a parameter to linearly connect $\sigma_E$ to $\Delta\Sigma$, with initial value 1 for proton and 10 for carbon. $R_E$ is adjusted according to the values of $\Delta\Sigma$ in the two previous steps. If $\Delta\Sigma_1 * \Delta\Sigma_2 > 0$ (where the indices 1 and 2 correspond to the two steps before the current step), $R_E$ is scaled by 1.25, otherwise, $R_E$ is divided by 1.25. We set the tolerance for $\Delta\Sigma/\Sigma_M$ as 1%.

## 3.2 Parameters tuning for Lateral dose profile spot sizes

Once the longitudinal parameters ($E_r$ and $\sigma_E$) are adjusted, we tune the lateral parameters by comparing lateral dose spot sizes. Spot sizes along the i (where i=x,y) axis at different depths are mainly determined by $\sigma_i$ and $\sigma_{\phi i}$, while the gradient of spot sizes at different positions is mainly determined by $\sigma_{\phi i}$. Therefore, we tune parameters for both x and y axes in the same step.

Similarly to the $\sigma_E$ tuning, we divide the process into two ranges depending on whether the simulated spot sizes are significantly larger than measured ones. Spot sizes at five different depths are provided, however we use the spot sizes at $z_{-2}$ and $z_0$ to define the ranges. If the difference between the widths of the calculated and measured distributions at $z_{-2}$ or $z_0$ is larger than 50%, the iteration is assigned to the first range. In this case, the width is modified according to the expression:

- $\sigma_i = \sigma_i\ (1 + 0.5\ \Delta\sigma_i(z_{-2})/|\Delta\sigma_i(z_{-2})|)$,
- $\sigma_{\phi i} = \sigma_{\phi i}\ (1 + 0.5\ \Delta\sigma_i(z_0)/|\Delta\sigma_i(z_0)|)$,

Where $\Delta\sigma_i(z)$ is the difference of the spot with measurement and calculations at position z mm. if the iteration results fall in this range, it means we selected too large values for $\sigma_i$ or $\sigma_{\phi i}$.

If the difference between the widths of the calculated and measured distributions at $z_{-2}$ or $z_0$ is less than 50%, simulations fall to the 2$^{nd}$ range. We tune the parameters by checking three criteria.

1. The spot size gradient as a function of z, which we evaluate by looking at the ratio of the spot widths at the shallowest ($z_{-2}$) and the deepest depths ($z_2$), $g_i = \sigma_i(z_2)/\sigma_i(z_{-2})$. We also define $\Delta g_i = g_i^m - g_i^c$, where $g_i^m$ and $g_i^c$ are the measured and calculated values of $g_i$ respectively. We tune $\sigma_{\phi i}$ with the function $\sigma_{\phi i} = \sigma_{\phi i} + \lambda_i \Delta g_i/g_i^m$ to make $g_i^c$ match $g_i^m$. The parameter $\lambda_i$ is set to initial value <u>0.01</u> for proton and 0.0025 for carbon. It is modified in each iteration according to how $\Delta g_i$ varies in the two previous steps. If $\Delta g_{i1} \Delta g_{i2} > 0$, $\lambda_i$ is scaled by 1.25. Otherwise $\lambda_i$ is divided by 1.25.
2. $\sigma_i$ is tuned with the function $\sigma_i = \sigma_i + \kappa_i \Delta\sigma_i(z_0)/\sigma_i^m(z_0)$. The parameter $\kappa_i$ is initially set to 0.25 for proton and 0.025 for carbon. It then varies according to changes of $\Delta\sigma_i(z_0)$ in the last two steps.
3. At the end we check if the spot size at $z_{-2}$ matches with the measured one. The method is the same as we tune the spot size at isocenter.

We set tolerances for the difference of spot sizes at $z_{-2}$ and $z_0$ for both x and y axes as 2.5% and 2% respectively. We also set tolerance for the gradient $g_i$ to be 3%. The maximum iterations for lateral parameters tuning is 50 for proton and 35 for carbon. If tolerances set for spot sizes or gradients can not be reached till the the maximum loop, we pick up the best result among all the iterations. We define the best result for an iteration as having smallest difference between simulated and measured spot sizes at 5 depths but with higher weight (1/6.0 for isocenter comparing with 1/12.0 for other depths) at isocenter.

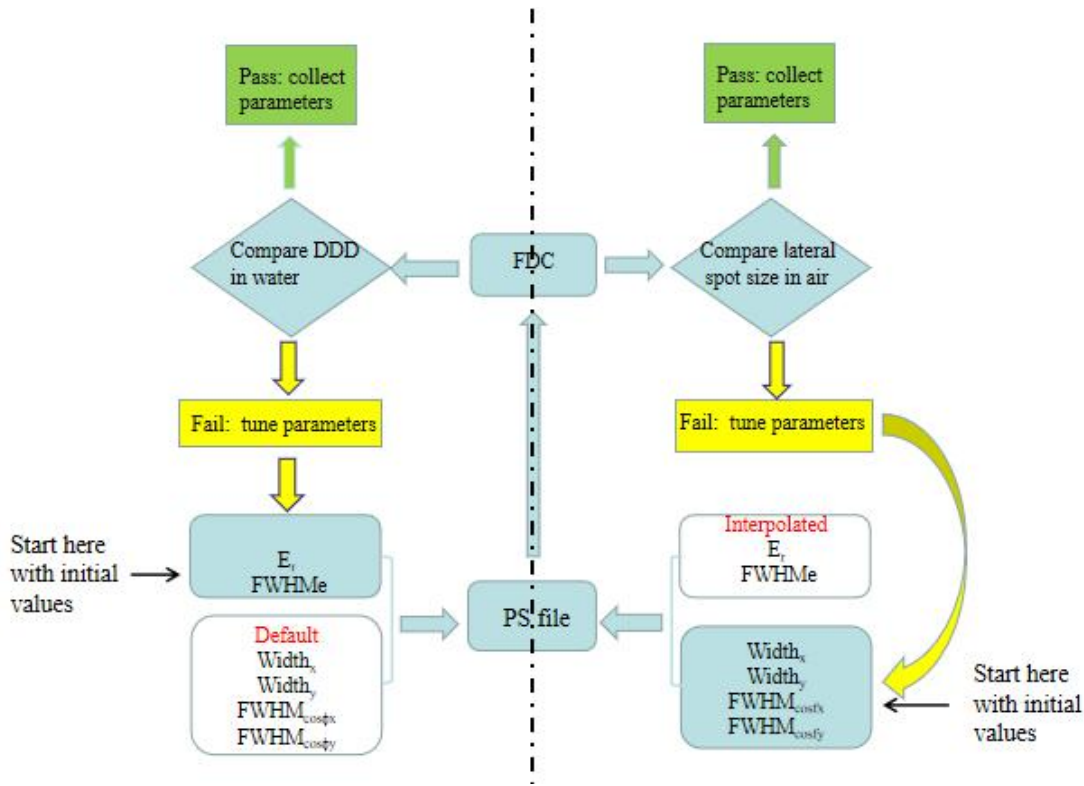

**Figure 1** Algorithm flow. The left side of dash line is for Step 1 tuning $E_r$ and $\sigma_E$ for DDD and the right side of the dash line is for Step 2 tuning $\sigma_x$, $\sigma_y$, $\sigma_{\phi x}$ and $\sigma_{\phi y}$ for lateral dose spot size.

## 4 Results:

We selected nine energies for proton and carbon respectively to automatically tune the parameters $E_r$ and $\sigma_E$ to make their simulated DDD in water match measurements. Results are presented in Figure 2 and 3. The two parameters for the rest of the energies can be linearly interpolated from these of the 9 energies.

For lateral dose spot size tuning with our automatic code, 14 energies were selected for each particle and focus. Parameters for other energies can be linearly interpolated from those of the selected 14 energies. Results are displayed in Figure 4 and 5 and Table 1 and 2.

### 4.1 Depth Dose distribution in water

Figure 1 displays DDD comparison between measurements and FDC simulations with PS files generated with automatically tuned parameters for 9 single proton energies. Figure 2 depicts the DDD comparison for 9 single carbon energies with a 3mm ripple filter. Simulations match measurements very well except for the proton energy 48.1 MeV. Nevertheless, the largest dose spatial shift in the plot for that worst case is around 0.5 mm, which is well within the clinical tolerance. With the criterion 0.4%/0.4 mm, gamma index passing rates (Low *et al* 1998, Clasie *et al* 2012) for proton energies 48.1, 85.7, 104.5, 124.2, 141.9, 163.3, 182.9, 199.0 and 221.1 MeV are 80.8%, 100.0%, 100.0%, 100.0%, 100.0%, 100.0%, 100.0%, 99.5% and 99.2% respectively. For carbon energies with a 3 mm ripple filter, gamma-index passing rates with the same criterion are 100% for 159.5, 195.7, 234.1, 268.9, 311.6, 351.4 MeV/u. And the passing rates are 96.5% for 88.7 MeV/u, 99.3% for 398.2 MeV/u and 97.1% for 430.1 MeV/u. If we lower the criterion to 0.6%/0.6 mm, the passing rate increases to 96.2% for the proton energy 48.1 MeV. When we use 1%/1 mm as the criterion, all the passing rates for different energies are 100.0%. These high passing rates with strict criteria confirms that simulations match well with measurements and confirm the automatic code presented here can find appropriate parameters for DDD.

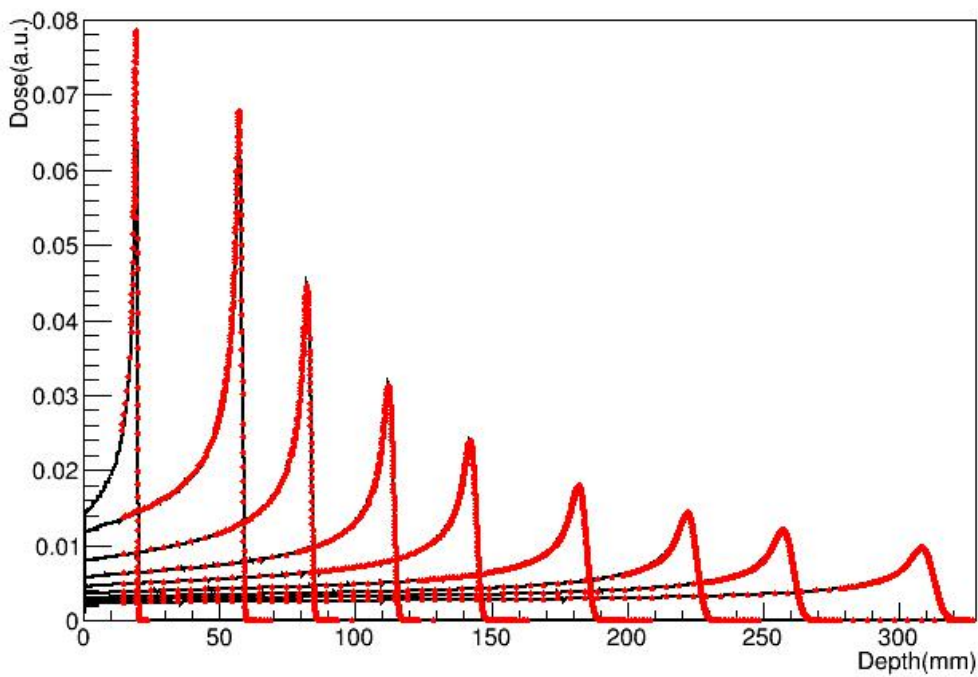

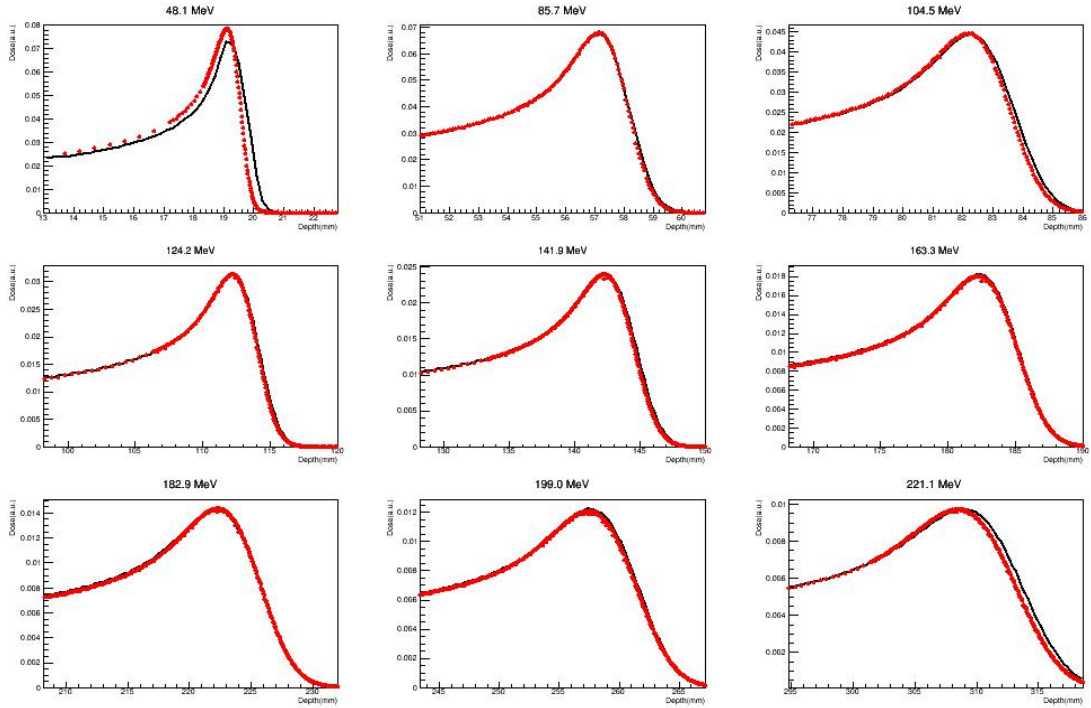

**Figure 2** Depth dose distributions in water for different proton single energies. Black curves are for FDC simulations and red ones are for measurements. Plots in the lower panel show the same content as those in the upper panel but with details around the Bragg peaks enlarged.

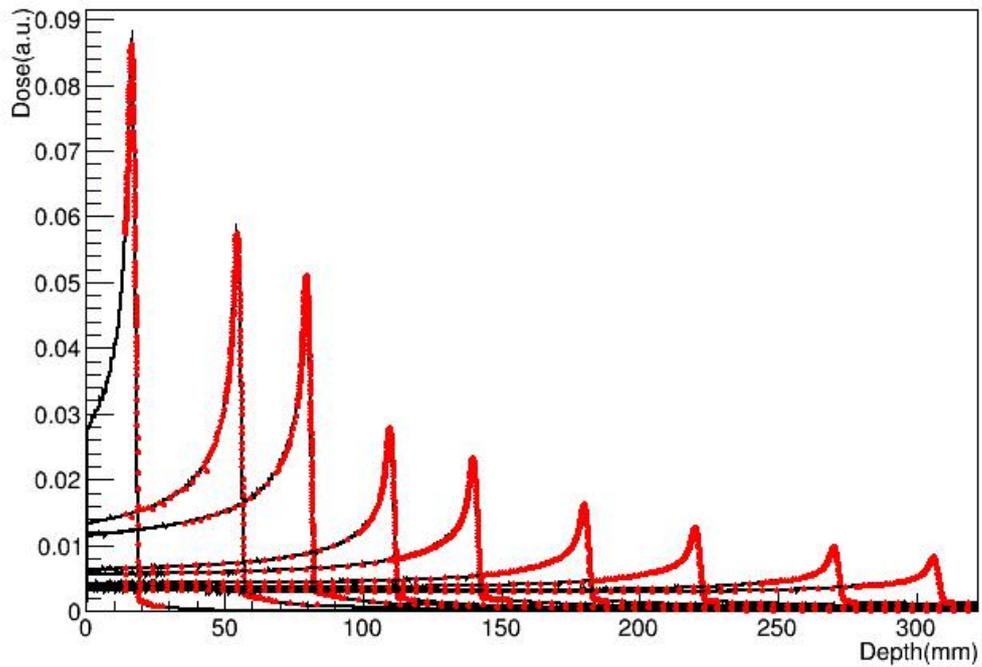

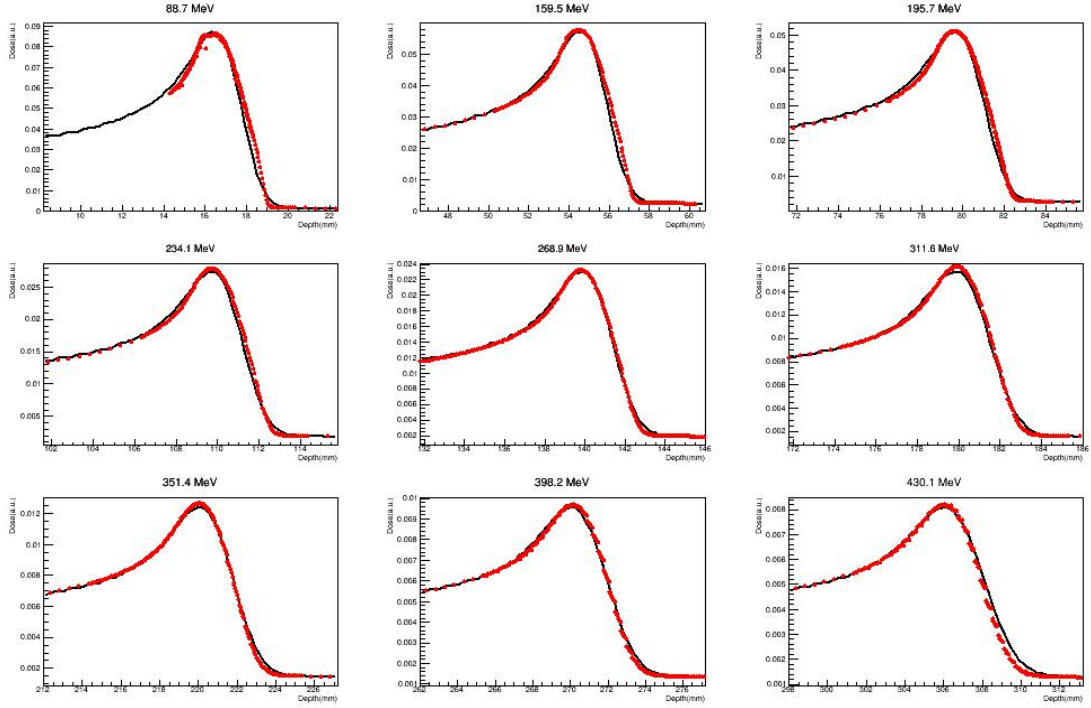

**Figure 3** Depth dose distributions in water for different carbon single energies. Black curves are for FDC simulations and red ones are for measurements. Plots in the lower panel show the same content as those in the upper panel but with details around peaks enlarged.

**4.2 Spot sizes of lateral dose profile**

We list measured ($M_i$) and simulated ($S_i$) spot sizes at different depths in Table 1 (for proton) and 2 (for carbon). In general, values in $S_i$ column agree well with those in $M_i$ in both tables. We calculated the deviation percentage of spot sizes between simulation and measurement averaged over five depths for both x and y axes for each energy with the formula: $\Delta = \frac{1}{5} \sum_{i=-2}^{2} |S_i - M_i| / M_i$. Values are listed in the last but one column of each table. Isocenter is usually where the patient located. Therefore, the agreement at depth 0 is the most relevant. We also listed the deviation percentage at isocenter ( $\Delta_{Iso} = |S_0 - M_0| / M_0$ ) in the last column in each table. The averaged deviations and deviations at isocenter for all tuned proton energies in Table 1 are less than 1.8% and 2.2%. For carbon in Table 2, these two values are less than 3.4% and 4.8%, which are higher than that for proton energies. However, except the two largest deviations 2.8% and 4.8%, the other values are no more than 2.3% and 1.6%.

In addition, we selected 5 energies for each particle to visualize their measured and simulated spot sizes at different depths as shown in Figure 4 (for proton) and 5 (for carbon). Both plots show good agreement between measurements and simulations.

| $E_n$ MeV | Axis | Spot size | | | | | | | | | | $\Delta$ % | $\Delta_{Iso}$ % |
|---|---|---|---|---|---|---|---|---|---|---|---|---|---|
| | | -800 mm | | -400 mm | | 0 mm | | 200 mm | | 600 mm | | | |
| | | $M_{-2}$ | $S_{-2}$ | $M_{-1}$ | $S_{-1}$ | $M_0$ | $S_0$ | $M_1$ | $S_1$ | $M_2$ | $S_2$ | | |
| 48.1 | x | 12.0 | 12.2 | 21.1 | 20.4 | 31.5 | 31.2 | 36.9 | 37.0 | 47.7 | 47.6 | 1.3 | 1.0 |
| | y | 13.1 | 13.2 | 21.4 | 21.8 | 31.5 | 31.4 | 36.7 | 37.2 | 47.3 | 46.6 | 1.2 | 0.3 |
| 59.8 | x | 11.6 | 11.4 | 18.1 | 17.8 | 26.0 | 26.4 | 30.3 | 29.8 | 38.9 | 39.2 | 1.5 | 1.5 |
| | y | 12.4 | 12.2 | 18.4 | 18.4 | 26.0 | 25.8 | 30.1 | 30.2 | 38.5 | 39.0 | 0.8 | 0.8 |
| 85.7 | x | 8.9 | 9.0 | 13.3 | 13.4 | 18.7 | 18.4 | 21.7 | 21.8 | 27.7 | 27.6 | 0.9 | 1.6 |
| | y | 10.7 | 10.8 | 14.5 | 14.6 | 19.4 | 19.6 | 22.0 | 22.4 | 27.7 | 28.2 | 1.3 | 1.0 |
| 104.5 | x | 8.8 | 8.6 | 11.9 | 12.2 | 16.0 | 16.0 | 18.5 | 18.4 | 23.4 | 23.6 | 1.2 | 0.0 |
| | y | 10.9 | 11.0 | 13.1 | 13.0 | 16.9 | 16.8 | 19.0 | 19.0 | 23.4 | 23.4 | 0.5 | 0.6 |
| 124.2 | x | 9.0 | 9.0 | 11.4 | 11.4 | 14.3 | 14.6 | 16.4 | 16.6 | 20.4 | 20.8 | 1.1 | 2.1 |
| | y | 10.8 | 11.0 | 12.6 | 12.8 | 15.2 | 15.0 | 16.9 | 16.8 | 20.5 | 20.6 | 1.2 | 1.3 |
| 136.2 | x | 9.2 | 9.0 | 11.2 | 11.0 | 13.6 | 13.4 | 15.5 | 15.4 | 19.1 | 19.0 | 1.3 | 1.5 |
| | y | 10.8 | 11.0 | 12.3 | 12.4 | 14.5 | 14.6 | 16.0 | 16.0 | 19.2 | 19.4 | 0.9 | 0.7 |
| 141.9 | x | 9.2 | 9.0 | 11.1 | 11.0 | 13.3 | 13.4 | 15.0 | 15.0 | 18.5 | 18.4 | 0.9 | 0.8 |
| | y | 10.9 | 11.0 | 12.2 | 12.4 | 14.2 | 14.2 | 15.6 | 15.4 | 18.6 | 18.4 | 1.0 | 0.0 |
| 150.1 | x | 9.3 | 9.2 | 10.9 | 10.6 | 12.9 | 13.0 | 14.5 | 14.2 | 17.8 | 17.4 | 1.8 | 0.8 |
| | y | 10.9 | 10.8 | 12.1 | 11.8 | 13.9 | 14.2 | 15.2 | 15.2 | 18.0 | 18.0 | 1.1 | 2.2 |
| 163.3 | x | 9.7 | 9.6 | 10.8 | 10.6 | 12.6 | 12.6 | 13.9 | 13.8 | 16.9 | 16.4 | 1.3 | 0.0 |
| | y | 10.9 | 10.8 | 11.7 | 11.6 | 13.4 | 13.4 | 14.5 | 14.4 | 17.0 | 16.8 | 0.7 | 0.0 |
| 175.7 | x | 9.2 | 9.2 | 10.4 | 10.2 | 11.8 | 11.6 | 13.1 | 13.0 | 15.9 | 15.4 | 1.5 | 1.7 |

| | | | | | | | | | | | | |
|---|---|---|---|---|---|---|---|---|---|---|---|---|
| | y | 10.7 | 10.8 | 11.5 | 11.6 | 12.9 | 13.0 | 13.9 | 13.8 | 16.1 | 16.0 | 0.8 | 0.8 |
| 182.9 | x | 8.8 | 8.8 | 9.9 | 9.8 | 11.4 | 11.6 | 12.6 | 12.6 | 15.2 | 15.4 | 0.8 | 1.8 |
| | y | 10.4 | 10.4 | 11.2 | 11.2 | 12.5 | 12.6 | 13.6 | 13.4 | 15.6 | 15.6 | 0.5 | 0.8 |
| 199.0 | x | 8.4 | 8.4 | 9.3 | 9.4 | 10.6 | 10.8 | 11.6 | 11.8 | 14.1 | 14.0 | 1.1 | 1.9 |
| | y | 10.1 | 10.0 | 10.8 | 10.6 | 11.9 | 11.8 | 12.8 | 12.8 | 14.7 | 14.4 | 1.1 | 0.8 |
| 205.7 | x | 8.4 | 8.2 | 9.1 | 8.8 | 10.4 | 10.2 | 11.4 | 11.4 | 13.8 | 13.6 | 1.8 | 1.9 |
| | y | 10.4 | 10.6 | 10.7 | 10.8 | 11.8 | 12.0 | 12.7 | 13.0 | 14.4 | 14.6 | 1.7 | 1.7 |
| 221.1 | x | 8.6 | 8.6 | 9.3 | 9.0 | 10.3 | 10.4 | 11.1 | 11.0 | 13.3 | 13.0 | 1.5 | 1.0 |
| | y | 10.8 | 10.8 | 11.2 | 11.2 | 12.0 | 11.8 | 12.7 | 12.4 | 14.2 | 13.8 | 1.4 | 1.7 |

**Table 1** Measured ($M_i$, where $i$ is from -2 to 2) and simulated ($S_i$, $i$ is from -2 to 2) spot sizes at 5 different depths for different proton energies and deviation percentage of simulated spot sizes from measured ones at isocenter ($\Delta_{Iso}$) and averaged over 5 depths ($\Delta$).

| $E_n$ MeV/u | Axis | Spot sizes | | | | | | | | | | $\Delta$ % | $\Delta_{Iso}$ % |
|---|---|---|---|---|---|---|---|---|---|---|---|---|---|
| | | -800 mm | | -400 mm | | 0 mm | | 200 mm | | 600 mm | | | |
| | | $M_{-2}$ | $S_{-2}$ | $M_{-1}$ | $S_{-1}$ | $M_0$ | $S_0$ | $M_1$ | $S_1$ | $M_2$ | $S_2$ | | |
| 86.2 | X | 7.0 | 6.8 | 9.3 | 9.4 | 12.7 | 12.6 | 14.5 | 14.6 | 18.4 | 18.6 | 1.3 | 0.8 |
| | Y | 7.4 | 7.4 | 9.5 | 9.8 | 12.8 | 13.0 | 14.6 | 14.8 | 18.3 | 18.2 | 1.3 | 1.6 |
| 110.6 | X | 7.2 | 7.2 | 8.6 | 8.8 | 10.9 | 10.8 | 12.2 | 12.2 | 15.0 | 15.4 | 1.2 | 0.9 |
| | Y | 7.3 | 7.2 | 8.7 | 8.8 | 10.8 | 10.8 | 12.1 | 12.0 | 14.9 | 14.8 | 0.8 | 0.0 |
| 159.5 | X | 6.8 | 6.8 | 7.4 | 7.2 | 8.7 | 8.6 | 9.5 | 9.4 | 11.2 | 11.4 | 1.3 | 1.1 |
| | Y | 6.9 | 6.8 | 7.6 | 7.6 | 8.8 | 8.8 | 9.6 | 9.6 | 11.3 | 11.2 | 0.5 | 0.0 |
| 195.7 | X | 6.7 | 6.6 | 7.1 | 6.8 | 7.9 | 7.8 | 8.5 | 8.6 | 9.8 | 9.8 | 1.6 | 1.3 |
| | Y | 6.7 | 6.6 | 7.1 | 7.2 | 7.9 | 8.0 | 8.5 | 8.4 | 9.8 | 9.6 | 1.5 | 1.3 |
| 227.9 | X | 6.4 | 6.4 | 6.7 | 6.8 | 7.3 | 7.4 | 7.7 | 7.8 | 8.8 | 8.8 | 0.8 | 1.4 |

| | | | | | | | | | | | | |
|---|---|---|---|---|---|---|---|---|---|---|---|---|
| | Y | 6.6 | 6.6 | 6.9 | 6.8 | 7.5 | 7.6 | 7.9 | 8.0 | 8.9 | 8.8 | 1.0 | 1.3 |
| 268.9 | X | 6.5 | 6.4 | 6.5 | 6.8 | 6.9 | 7.0 | 7.2 | 7.4 | 7.9 | 8.0 | 2.3 | 1.4 |
| | Y | 6.6 | 6.6 | 6.7 | 6.8 | 7.2 | 7.2 | 7.5 | 7.6 | 8.3 | 8.4 | 0.8 | 0.0 |
| 285.4 | X | 6.4 | 6.4 | 6.4 | 6.6 | 6.7 | 6.8 | 7.0 | 7.2 | 7.6 | 7.8 | 2.0 | 1.5 |
| | Y | 6.6 | 6.6 | 6.7 | 6.8 | 7.1 | 7.2 | 7.4 | 7.6 | 8.1 | 8.2 | 1.4 | 1.4 |
| 311.6 | X | 6.5 | 6.4 | 6.5 | 6.6 | 6.7 | 6.6 | 6.9 | 6.8 | 7.4 | 7.2 | 1.7 | 1.5 |
| | Y | 6.7 | 6.8 | 6.8 | 6.8 | 7.0 | 7.0 | 7.3 | 7.2 | 7.8 | 7.8 | 0.6 | 0.0 |
| 336.8 | X | 6.5 | 6.4 | 6.4 | 6.4 | 6.4 | 6.4 | 6.6 | 6.6 | 7.0 | 6.8 | 0.9 | 0.0 |
| | Y | 6.8 | 6.8 | 6.8 | 6.8 | 7.0 | 7.0 | 7.2 | 7.2 | 7.7 | 7.6 | 0.3 | 0.0 |
| 351.4 | X | 6.5 | 6.2 | 6.3 | 6.2 | 6.3 | 6.4 | 6.4 | 6.4 | 6.8 | 6.6 | 2.1 | 1.6 |
| | Y | 6.8 | 6.8 | 6.8 | 6.8 | 7.0 | 7.0 | 7.2 | 7.2 | 7.6 | 7.6 | 0.0 | 0.0 |
| 384.4 | X | 6.6 | 6.4 | 6.4 | 6.4 | 6.3 | 6.6 | 6.4 | 6.6 | 6.6 | 6.8 | 2.8 | 4.8 |
| | Y | 6.6 | 6.6 | 6.6 | 6.4 | 6.7 | 6.8 | 6.9 | 6.8 | 7.3 | 7.2 | 1.5 | 1.5 |
| 398.2 | X | 6.4 | 6.2 | 6.2 | 6.2 | 6.1 | 6.2 | 6.2 | 6.2 | 6.4 | 6.4 | 1.0 | 1.6 |
| | Y | 6.6 | 6.6 | 6.6 | 6.4 | 6.7 | 6.8 | 6.8 | 6.8 | 7.2 | 7.2 | 0.9 | 1.5 |
| 430.1 | X | 6.4 | 6.2 | 6.1 | 6.2 | 6.1 | 6.2 | 6.2 | 6.2 | 6.3 | 6.4 | 1.6 | 1.6 |
| | Y | 6.7 | 6.6 | 6.7 | 6.6 | 6.7 | 6.8 | 6.8 | 6.8 | 7.1 | 7.2 | 1.2 | 1.5 |

**Table 2** Measured ($M_i$, where $i$ is from -2 to 2) and simulated ($S_i$, where $i$ is from -2 to 2) spot sizes at 5 different depths for different carbon energies with a 3 mm ripple filter and deviation percentage of simulated spot sizes from measured ones at isocenter ($\Delta_{Iso}$) and averaged over 5 depths ($\Delta$).

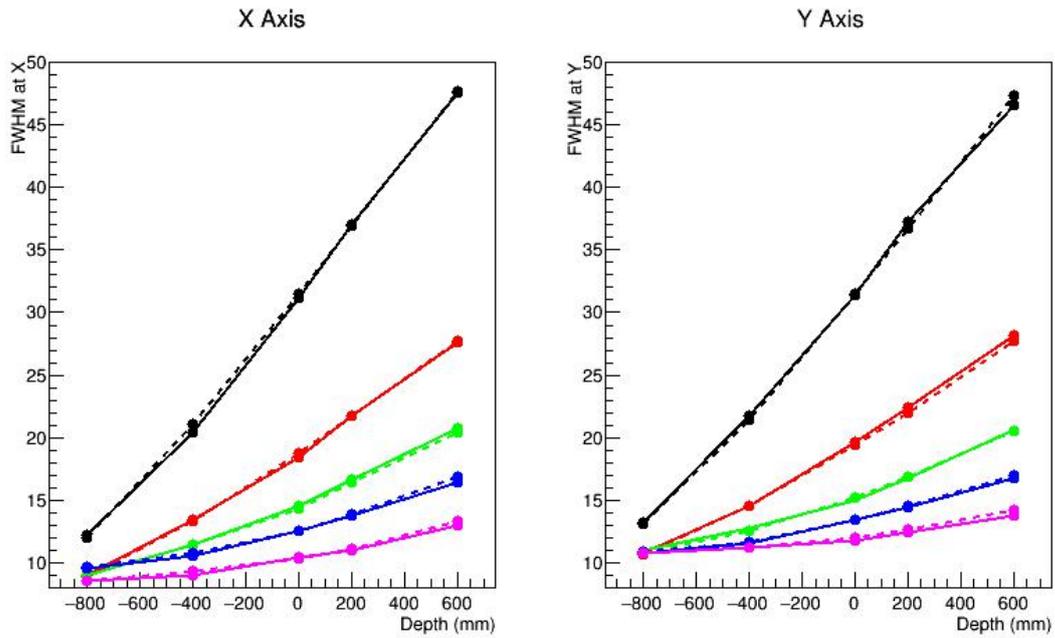

**Figure 4** Spot sizes at 5 different depths (-800, -400, 0, 200 and 600 mm) for five proton energies: 48.1 MeV (black), 85.7 MeV (read),124.2 MeV (green), 163.3 MeV (blue) and 221.1 MeV (pink). Solid lines with points are for simulation and solid lines with points are for measurements.

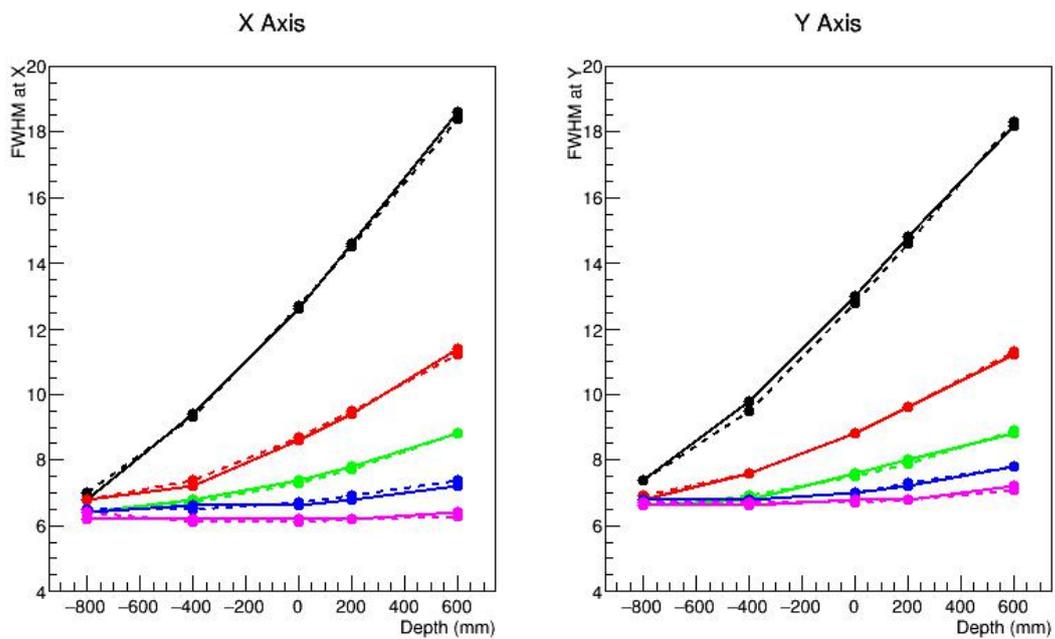

**Figure 5** Spot sizes at 5 different depths (-800, -400, 0, 200 and 600 mm) for five carbon energies with 3 mm ripple filter: 86.2 MeV/u (black), 159.5 MeV/u (red), 227.9 MeV/u (green), 311.6 MeV/u (blue) and 430.1 MeV/u (pink). Solid lines with points are for simulation and dash lines with points are for measurements.

### 4.3 Time

The average time to finish tuning of the longitudinal parameters is approximately 1 minute per proton energy and 10.5 minutes per carbon energy when the dose calculation is performed using 1 GPU. The average time to finish the tuning of the lateral parameters is about 8 minutes per proton energy and 9 minutes per carbon energy using 1 GPU.

### 5 Summary

We introduced an automatic algorithm to determine parameters for phase space files to be used in the Monte Carlo dose calculations for particle therapy. To demonstrate the validation of this algorithm, we presented the comparison of depth dose distribution (DDD) in water and spot size of lateral dose profile between measurement and FDC simulation with phase space files generated with the automatic algorithm. For proton, the gamma-index passing rate with criterion 0.6%/0.6 mm for DDD comparison is above 96% for the selected 9 energies. The gamma-index passing rate of comparison with criterion 0.4%/0.4 mm for 9 selected carbon energies with 3 mm ripple filter are above 97%. If 1% /1 mm are used, passing rates are 100% for all energies discussed in this paper. The mean percent deviation of simulated spot size from measurements averaged over 5 depths in x or y axis for proton is no more than 1.8% for 14 proton selected energies. For selected carbon energies, the mean percent deviation is below 3.5%. The time used for tuning parameters for DDD comparison with this automatic algorithm is about 1 minute per proton energy and 10 minutes per carbon energy with 1 GPU. For tuning parameters for spot sizes comparison, it takes about 7 minutes for this algorithm to finish 1 proton energy and 12 minutes for 1 carbon energy with 1 GPU.

The high gamma-index passing rates with strict criteria when comparing simulated DDDs with measurements and low deviations of simulated spot sizes from measurements confirm the validation of our automatic algorithm. The speed of this algorithm in determining parameters for both proton and carbon energy also demonstrate its efficiency.